\documentclass{PoS}

\title{Generalized parton distributions from neutrino experiments}

\ShortTitle{Generalized parton distributions from neutrino experiments}

\author{B. Z. Kopeliovich\\
        Departamento de F\'{i}sica, Universidad T\'ecnica Federico Santa Mar\'{i}a, Casilla 110-V, Valpara\'{i}so,
Chile\\
        E-mail: \email{Boris.Kopeliovich@usm.cl}}

\author{Iv\'an Schmidt\\
        Departamento de F\'{i}sica, Universidad T\'ecnica Federico Santa Mar\'{i}a, Casilla 110-V, Valpara\'{i}so,
Chile\\
        E-mail: \email{Ivan.Schmidt@usm.cl}}

\author{\speaker{Marat Siddikov}\\
        Departamento de F\'{i}sica, Universidad T\'ecnica Federico Santa Mar\'{i}a, Casilla 110-V, Valpara\'{i}so,
Chile\\
        E-mail: \email{Marat.Siddikov@usm.cl}}

\abstract{The analysis of deeply virtual meson production is extended to neutrino-production of the pseudo-Goldstone mesons (pions, kaons, eta-mesons) on nucleons, with the flavor content of the recoil baryon either preserved, or changed to a hyperon from the same $SU(3)$ octet. We rely on the $SU(3)$ relations and express all the cross-sections in terms of the proton generalized parton distributions (GPDs). The corresponding amplitudes are calculated at the leading twist level and in the leading order in $\alpha_{s}$, using a phenomenological parametrization of GPDs. We also included in the analysis the electromagnetic $\mathcal{O}(\alpha_{em})$-corrections to neutrino-induced deeply virtual meson production ($\nu$DVMP). We found that such electromagnetic corrections decrease with $Q^{2}$ in the Bjorken regime less than the standard $\nu$DVMP handbag contribution, so the electromagnetic mechanism dominates  at large $Q^{2}$. The electromagnetic corrections give rise to an angular correlation between the lepton and hadron scattering planes with harmonics sensitive to the real and imaginary parts of the DVMP amplitude. These corrections constitute a few percent effect in the kinematics of the forthcoming \textsc{MINERvA} experiment at Fermilab and should be taken into account in precise tests of GPD parametrizations. For virtualities $Q^{2}\sim 100$ GeV$^{2}$ these corrections become on a par with $\nu$DVMP handbag contributions.}

\FullConference{XXI International Workshop on Deep-Inelastic Scattering and Related Subject -DIS2013,\\
		22-26 April 2013\\
		Marseilles, France}

\begin{document}

\section{Introduction}

Nowadays one of the key objects used to parametrize nonperturbative
structure of the target are the generalized parton distributions (GPDs).
For the processes where the collinear factorization is applicable~\cite{JiCollins},
knowledge of the GPDs allows evaluation of the cross-sections for a wide class of processes.
Currently all information on GPDs comes from the electron-proton and
positron-proton processes measured at JLAB and HERA, in particular
from deeply virtual Compton scattering (DVCS) and deeply virtual meson
production 
(DVMP)~\cite{JiCollins,DVCSMPAll}.
The planned CLAS12 upgrade at JLAB~\cite{Kubarovsky:2011zz} will help to improve our understanding
of the GPDs even further. 

However, in practice extraction of GPDs from experimental data suffers from uncertainties caused by contributions of higher-twist components
of GPDs and pion distribution amplitudes (DAs) for moderate-$Q^{2}$ in JLAB kinematics~\cite{Ahmad:2008hp,Goldstein:2012az,Goloskokov:all},
uncertainties in vector meson DAs, which were never challenged experimentally.
In the kinematics of HERA the results of analyses can be affected by large BFKL-type logarithms in next-to-leading order (NLO) corrections~\cite{Ivanov:2007je}.

From this point of view, consistency checks of GPD extraction from
JLAB data, especially of their flavor structure, are important. Neutrino
experiments present a powerful tool, which could be used for this
purpose. The study of various processes in the Bjorken regime may be done
with the high-intensity \textsc{NuMI} beam at Fermilab, which 
will switch soon to the middle-energy (ME) regime with an
average neutrino energy of about $6$~GeV, and potentially can reach
energies up to 20 GeV, without essential loss of luminosity. With such a
setup the \textsc{MINERvA} experiment\textsc{~\cite{Drakoulakos:2004gn}}
should be able to probe the quark flavor structure of the targets.


Recently we discussed a possibility of extraction of the GPDs from the processes of deeply
virtual neutrinoproduction of the pseudo-Goldstone mesons ($\pi,\, K,\,\eta$)
on nucleons~\cite{Kopeliovich:2012dr}. The $\nu$DVMP measurements
with neutrino and antineutrino beams are complementary to the electromagnetic
DVMP. In the axial channel, due to the chiral symmetry breaking we
have an octet of pseudo-Goldstone bosons which act as a natural probe
of the flavor content. Due to the $V-A$ structure of the charged
current, in $\nu$DVMP one can access simultaneously the unpolarized
GPDs, $H,\, E$, and the helicity flip GPDs, $\tilde{H}$ and $\tilde{E}$.
As a consequence, we expect that the contributions of GPDs $H_{T},\, E_{T},\,\tilde{H}_{T},\,\tilde{E}_{T}$
which contribute multiplied by moments of the poorly known twist-3 pion DA $\phi_{p}$
should be negligibly small. Besides, important information on flavor structure
can be obtained by studying the transitional GPDs in the processes
with nucleon to hyperon transitions. As was discussed in~\cite{Frankfurt:1999fp},
assuming $SU(3)$ flavor symmetry, one can relate these GPDs  to the
ordinary diagonal GPDs in the proton.

\begin{figure}[htp]
\includegraphics[scale=0.35]{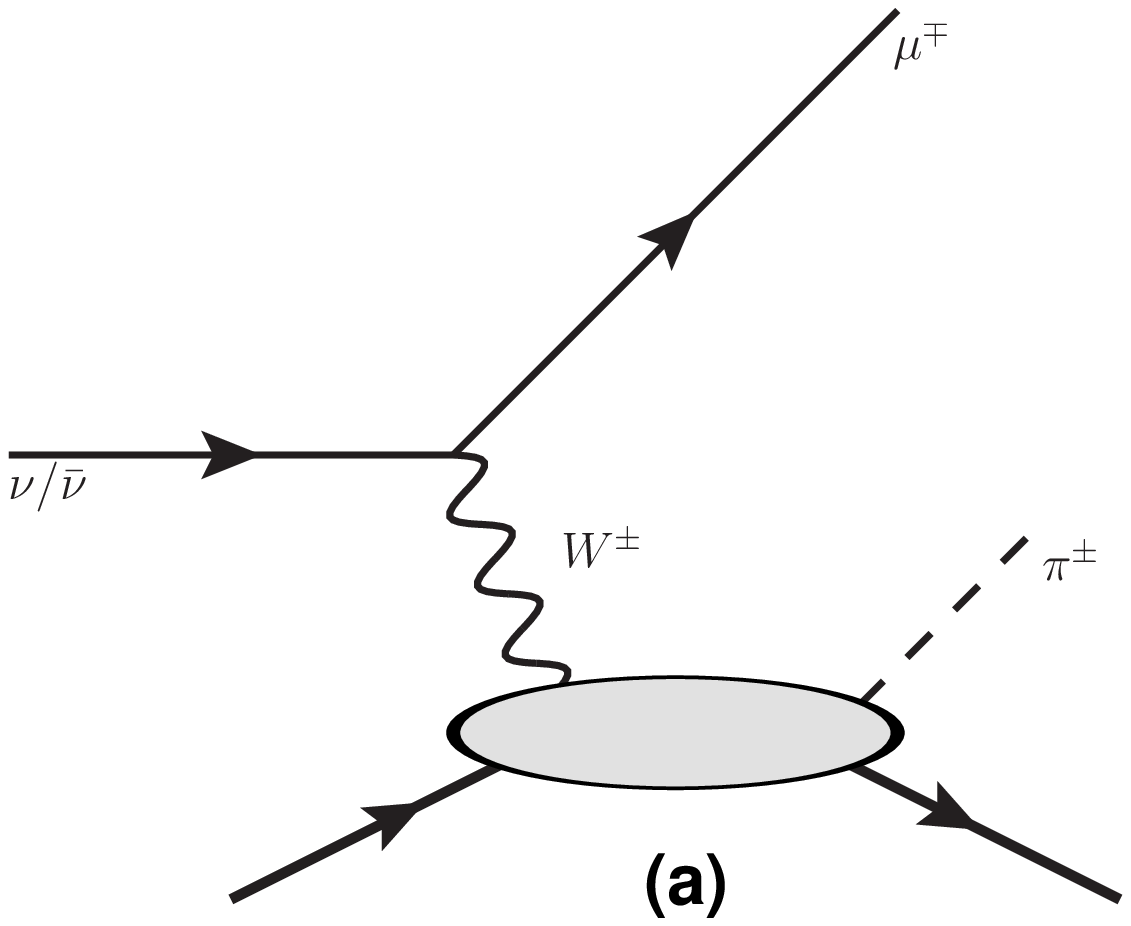}\includegraphics[scale=0.35]{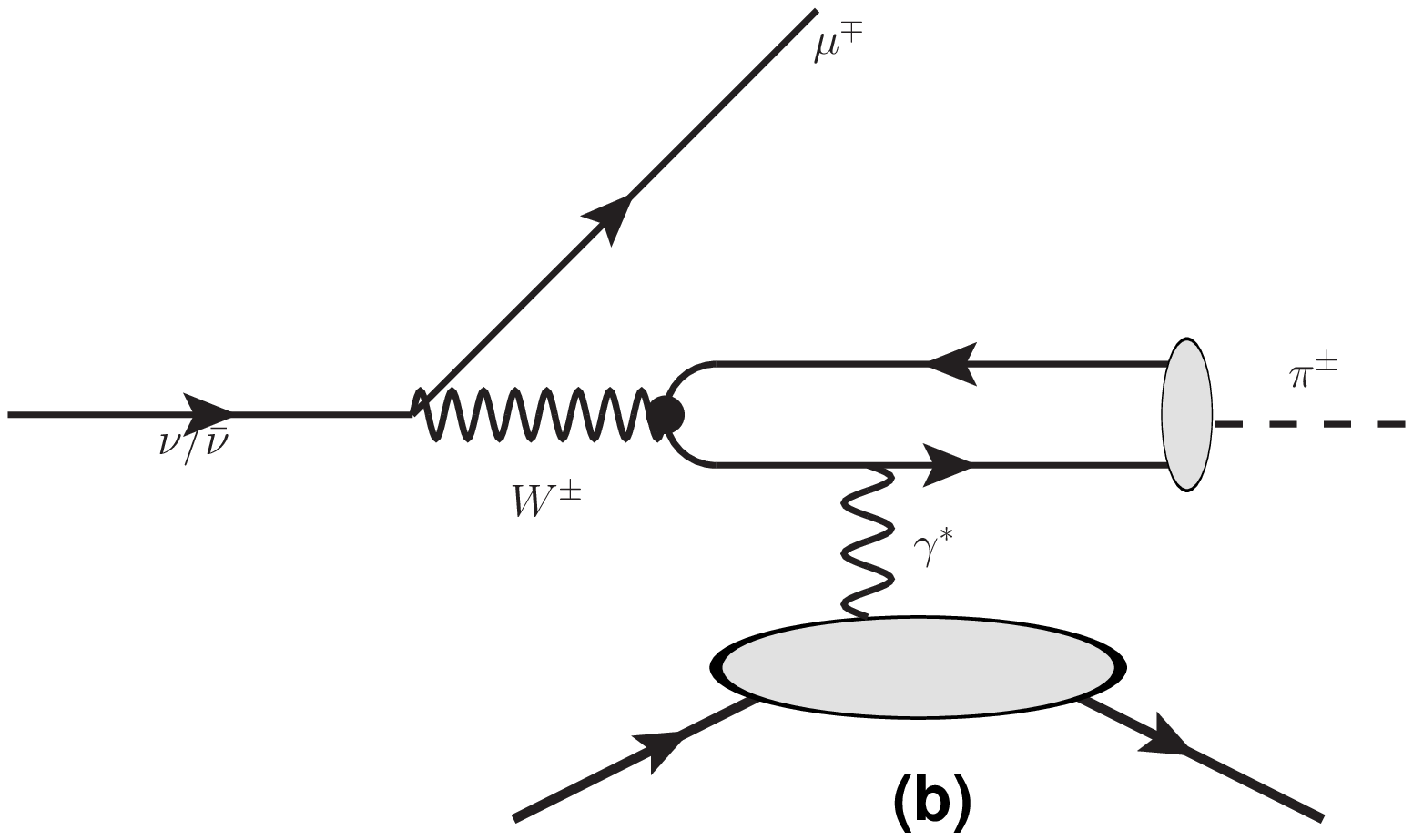}\includegraphics[scale=0.35]{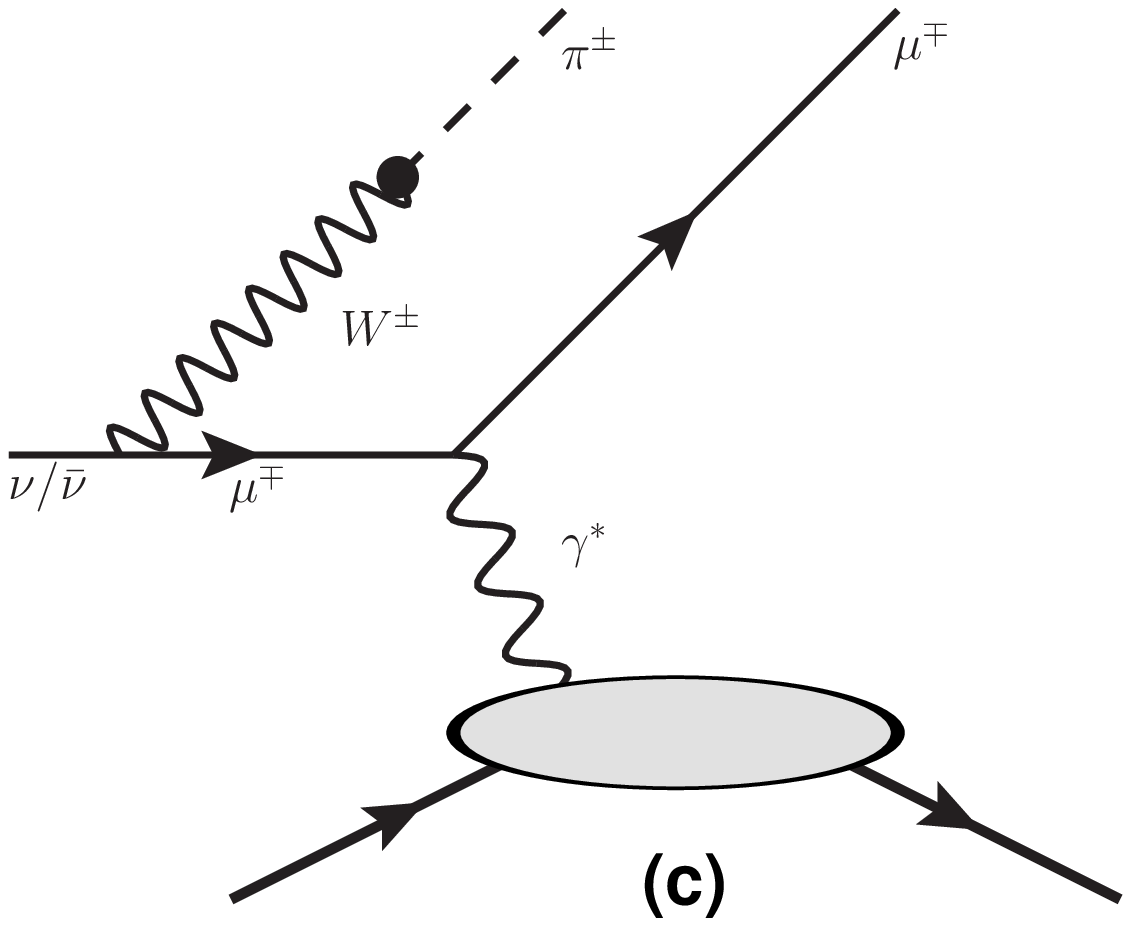}\caption{\label{fig:DVMPLT}Diagrams contributing to the neutrinoproduction
of mesons. (a) DVMP process (b,c) Electromagnetic corrections.}
\end{figure}

As was discussed in our recent paper~\cite{Kopeliovich:2013ae}, in analysis of the $\nu$DVMP one should include certain
electromagnetic contributions, in which the hadronic target interacts only by a single photon exchange, as shown schematically in Figure~(\ref{fig:DVMPLT}). 
Such diagrams resemble the Bethe-Heitler type contributions in Deeply Virtual Compton Scattering (DVCS). In $\nu$DVMP such processes are formally suppressed by $\alpha_{em}$, and for virtualities $Q^{2}\lesssim10$ GeV$^{2}$ relevant for modern
neutrino experiments they lead to few percent corrections.
However, in the Bjorken limit $Q^{2}\to\infty$ these contributions decrease with $Q^2$ less rapidly  than the $\nu$DVMP cross-section, and
already at $Q^{2}\sim100$~GeV$^{2}$ become comparable to the $\nu$DVMP result.
Existence of such diagrams opens a possibility to probe separately
the real and imaginary parts of the DVMP amplitude, in complete analogy to DVCS studies~\cite{Belitsky:all}.

\section{Numerical results and discussion}

\label{sec:Results}

\begin{figure}
\includegraphics[scale=0.4]{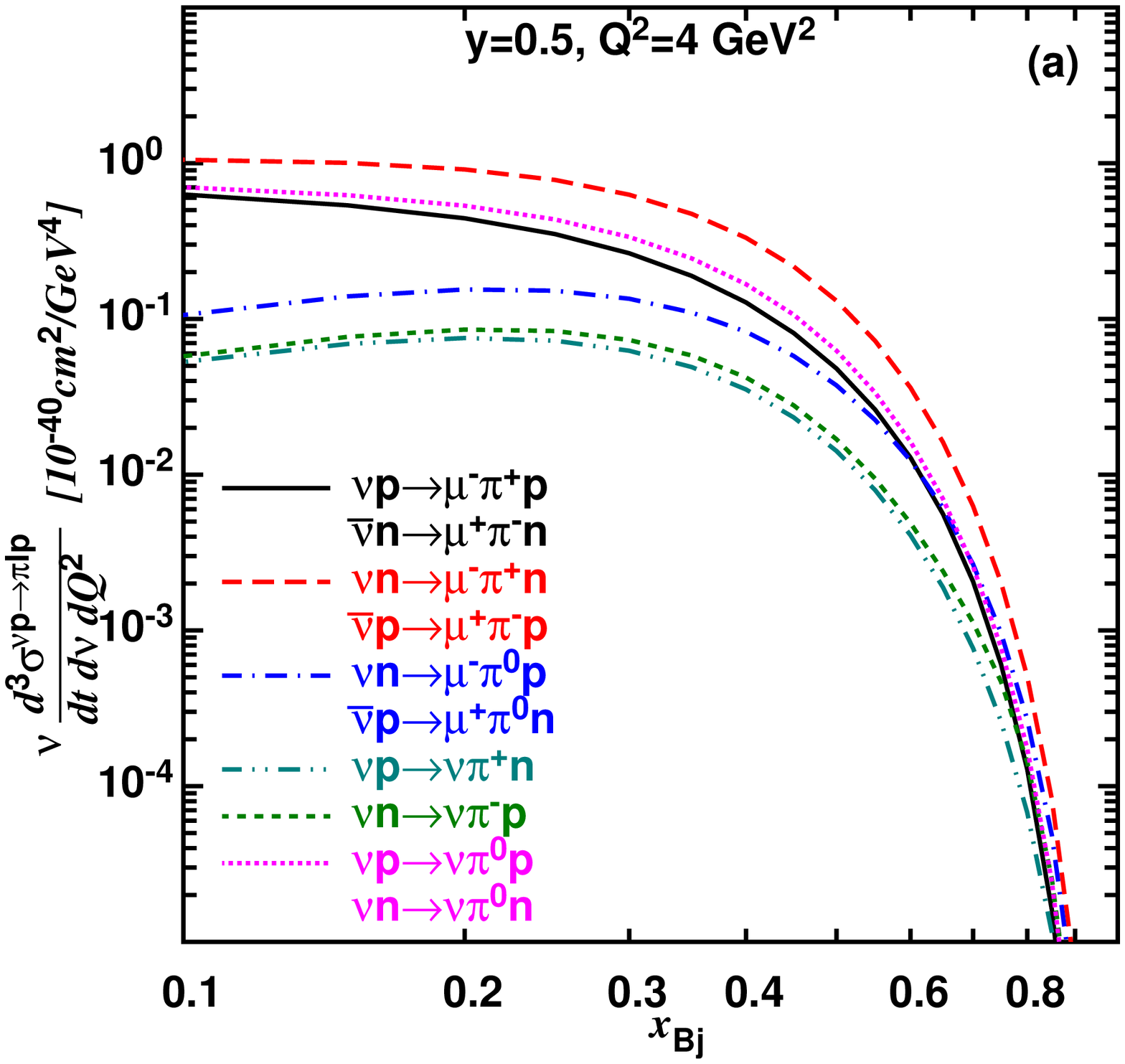}\qquad{}\includegraphics[scale=0.4]{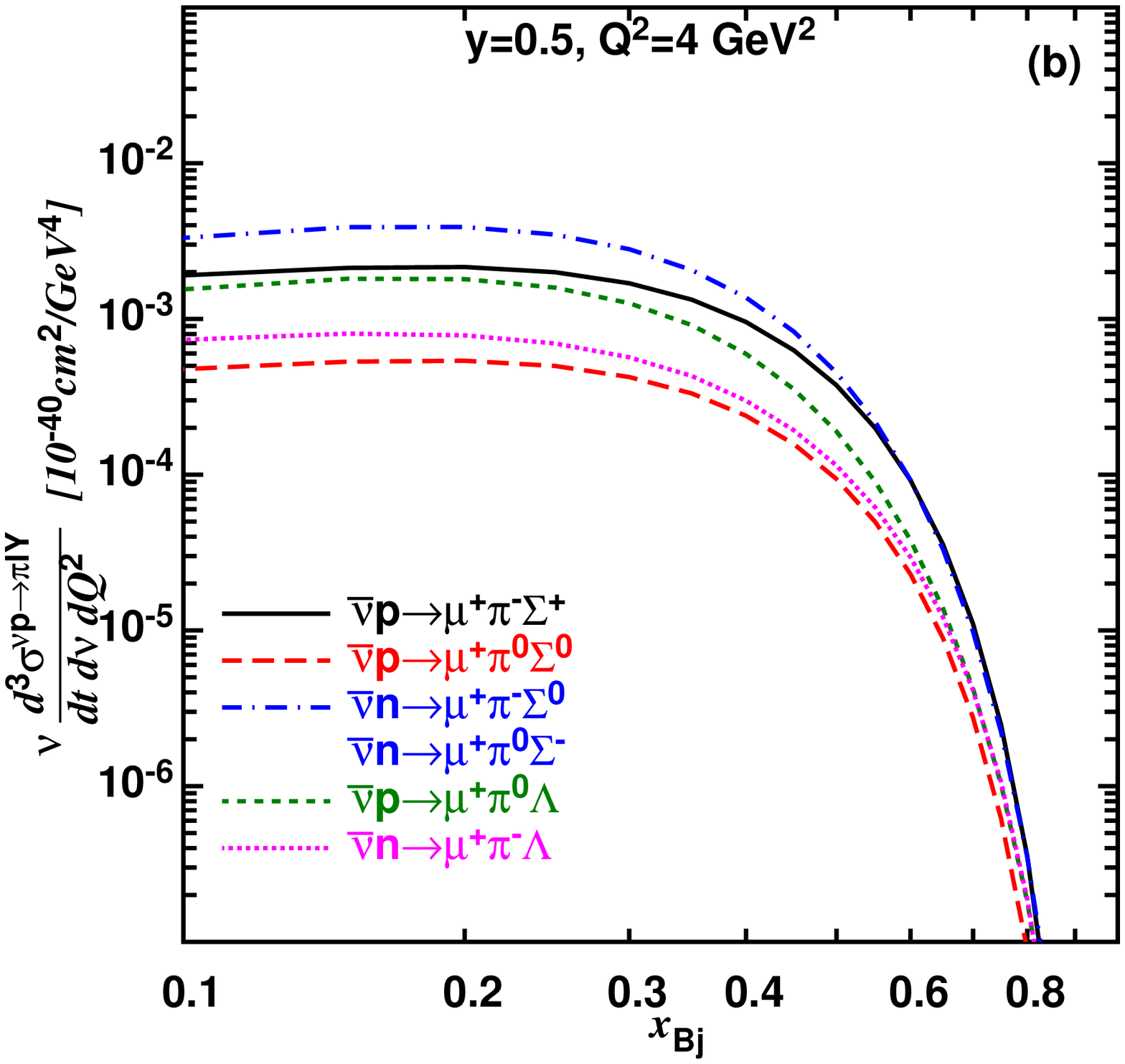}
\caption{\label{fig:DVMP-pions}(color online) Pion production on nucleons.
(a) Processes without strangeness production. (b) Processes with nucleon
to hyperon transition ($\Delta S=1$). Kinematics $t=t_{min}$ ($\Delta_{\perp}=0$)
is assumed.}
\end{figure}

In this section we present the results for the pion production calculated within the framework explained in detail in~\cite{Kopeliovich:2012dr}. For numerical estimates, we used the Kroll-Goloskokov parametrization~\footnote{The code for evaluation with arbitrary GPD parametrizations is available from ``\textit{Supplementary Material}'' of~\cite{Kopeliovich:2012dr,Kopeliovich:2013ae}} of 
GPDs~\cite{Goloskokov:all}.  The results for neutrino-production
of pions on nucleons are depicted in Figure~\ref{fig:DVMP-pions}.

The left pane of the Figure~\ref{fig:DVMP-pions} presents
the pion production processes without excitation of strangeness. 
For the diagonal channels, $p\to p$ and $n\to n$, we found that the production
rate of $\pi^{+}$ on neutrons is about twice as large as on protons.
This feature originates from the fact that the handbag diagram in the proton
probes the GPD $H_{d}$, whereas the larger $H_{u}$ contributes via the crossed
handbag; in the case of neutron they get swapped. At large $x_{Bj}\gtrsim0.6$
the corresponding cross-section is suppressed due to increase of $|t_{min}\left(x,Q^{2}\right)|$.

The off-diagonal processes with $p\rightleftarrows n$ transitions
are suppressed at small $x_{Bj}$ because they probe the GPD difference
$H_{u}-H_{d}$. In the small-$x_{Bj}$ regime ($x_{Bj}\lesssim0.1$)
the density of light sea quarks are expected to be equal, $\bar{d}\approx\bar{u}$,
and should cancel. The valence quark PDFs and the invariant amplitude $T_{M}$
behave like $\sim1/\sqrt{x_{Bj}},$ so that the cross-section vanishes
as $\sim x_{Bj}$. 
The cross-sections of the neutral $\pi^{0}$ production on the proton
and neutron (processes $\nu p\to\nu\pi^{0}p$ and $\nu n\to\nu\pi^{0}n$)
coincide under the assumption of $H$-dominance, otherwise they differ due to nonzero difference $\sim\tilde{H}_{u}-\tilde{H}_{d}$. Numerically
these effects are of order $1\,\%$ for the considered parametrization of GPDs, so the difference between the two curves is invisible in
the plot. A similar result holds for the processes $\nu p\to\nu\pi^{+}n$
and $\nu n\to\nu\pi^{-}p$: the corresponding cross-sections are equal
 under the assumption of $H$-dominance, otherwise they are different. As in the process of neutral pion production, the difference is controlled
by a small $\sim\tilde{H}_{u}-\tilde{H}_{d}$, however, due to suppression
of the GPD $H$ at small-$x_{Bj}$, those effects, which are proportional
$\sim\tilde{H}_{u}-\tilde{H}_{d}$ may be relatively large, so the difference
between the two cross-sections becomes visible in the plot.

In the right pane of Figure~\ref{fig:DVMP-pions} we show the cross-sections of pion production with
nucleon to hyperon transition. These cross-sections are Cabibbo-suppressed
and are hardly measurable in the \textsc{MINERvA} experiment. In contrast
to the $p\rightleftarrows n$ processes, at small $x_{Bj}$ the sea contributes
to the difference $H^{u}-H^{s},\, H^{d}-H^{s}$.
First of all, the sea flavor asymmetry appears due to the presence of
proton nonperturbative Fock components, like $p\to K\Lambda$.
This asymmetry vanishes in the invariant amplitude at small $x_{Bj}$
as $x_{Bj}^{-\alpha_{K^{*}}}$, where the intercept of the $K^{*}$
Reggeon trajectory is $\alpha_{K^{*}}\approx0.25$. Correspondingly,
this contribution to the cross section is suppressed as $\sim x_{Bj}^{1.5}$.

For kaon and $\eta$-meson production results are very similar and can be found in~\cite{Kopeliovich:2012dr}.

Now we would like to discuss the electromagnetic (EM) corrections shown in diagrams (b, c) in Figure~\ref{fig:DVMPLT}. These corrections are important only for the charged current induced meson production, when the internal state of the nucleon remains intact.

\begin{figure}
\includegraphics[scale=0.4]{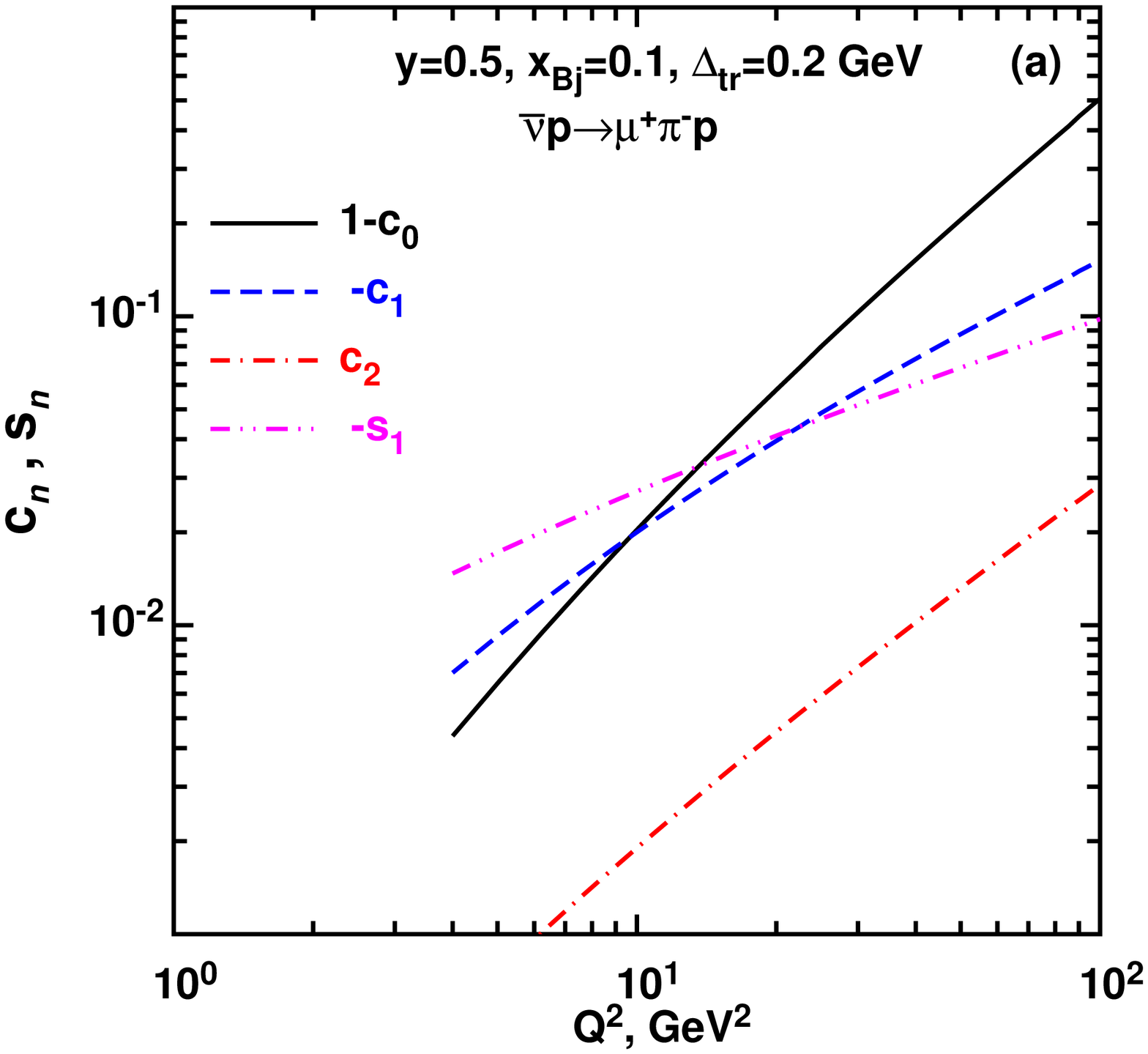}\qquad{}\includegraphics[scale=0.4]{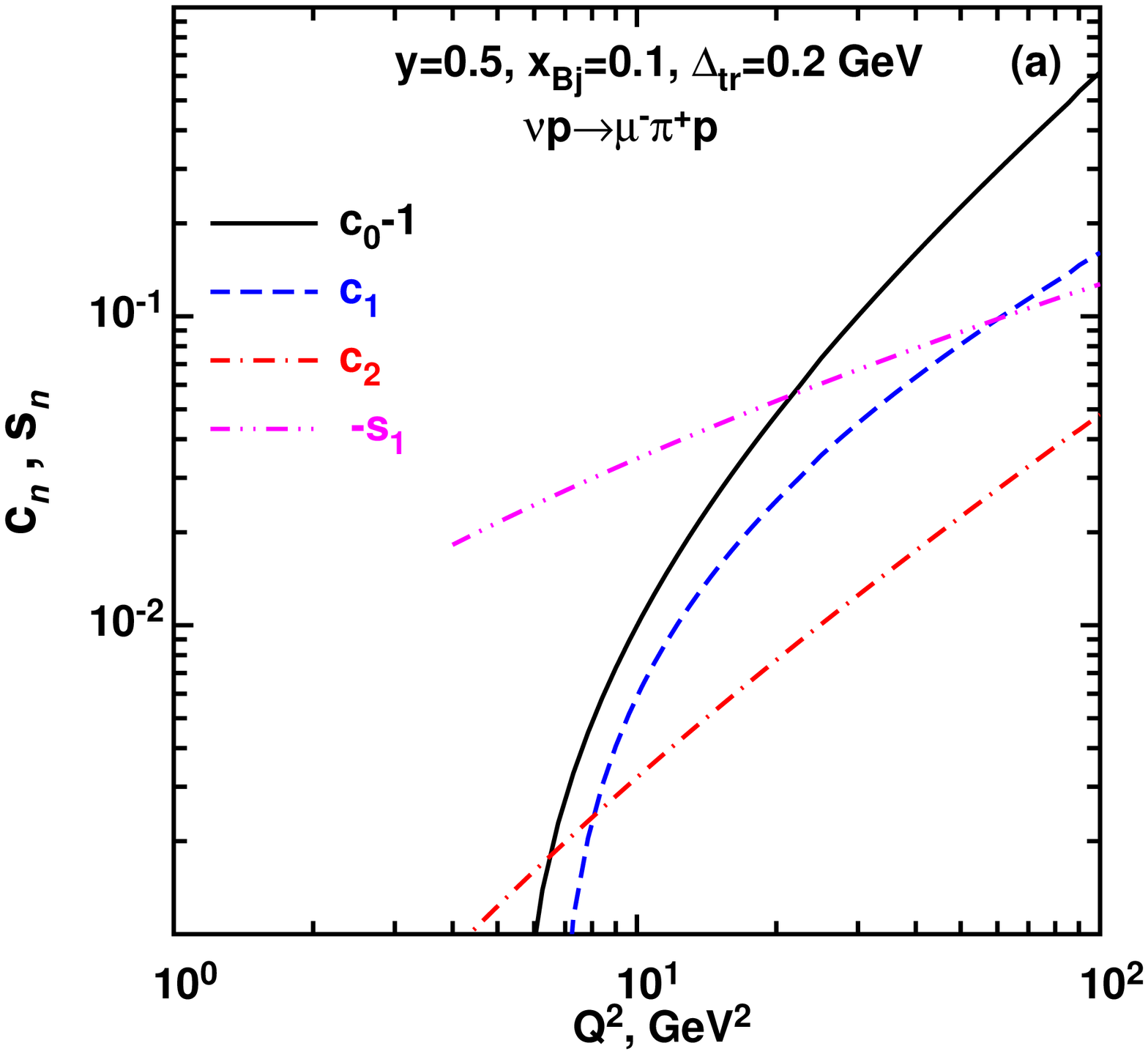}\\

\caption{\label{fig:DVMP-Q2}(color online) $Q^{2}$-dependence of the electromagnetic correction
to the $\nu$DVMP process. }
\end{figure}

 In the small-$Q^{2}$ regime the cross-section is dominated by the angular-independent DVMP
contribution, so the angular harmonics are small. For this reason
it is convenient to normalize all the coefficients to the DVMP cross-section,
namely
\begin{equation}
\frac{d^{4}\sigma}{dt\, d\ln x_{B}\, dQ^{2}d\phi}=\frac{d^{4}\sigma^{(DVMP)}}{dt\, d\ln x_{B}\, dQ^{2}d\phi}\left(\sum_{n=1}^{2}c_{n}\cos n\phi+s_{1}\sin\phi\right).
\end{equation}
 In the limit $\alpha_{em}\to0$ the coefficient $c_{0}=1$, all the
other coefficients vanish. The results for the $Q^{2}$-dependence
of the neutrino-production of pions and kaons on nucleons are depicted
in Figure~\ref{fig:DVMP-Q2}. As one can see, at small $Q^{2}\lesssim10$ GeV$^{2}$, for $\pi^{-}$
production all the coefficients $c_{n},\, s_{n}$ are of order a few 
percent, however they increase
rapidly with $Q^{2}$. The steepest growth has $1-c_{0}\sim Q^{2}$ modulo logarithmic corrections.
As one can see from the plot, due to the electromagnetic corrections 
the total cross-section at $Q^{2}\sim100$~GeV$^{2}$ is reduced to a half. The
asymmetry $s_{1}$ rises as a function of $Q^{2}$ and reaches
$\sim15\,\%$ at $Q^{2}=100$~GeV$^{2}$. 

The terms $c_{0}-1$ and $c_{1}$ get dominant contributions from
the interference term, for this reason they have different signs for
$\pi^{+}$ and $\pi^{-}$. The term $c_{2}$
gets contribution only from EM term, so it always has the same sign.
The $s_{1}$-term doesn't change sign under $C$-conjugation in lepton
part since it comes from $P$-odd interference between vector and
axial vector terms.

\section*{Acknowledgments}

This work was supported in part by Fondecyt (Chile) grants No. 1130543,
1100287 and 1120920. 


  \end{document}